\documentclass[12pt]{spieman}  
\usepackage{amsmath,amsfonts,amssymb}
\usepackage{graphicx}
\usepackage{setspace}
\usepackage{tocloft}
\usepackage[modulo]{lineno}
\usepackage{soul}

\renewcommand\hl[1]{#1}

\title{Improving the energy resolution of photon counting Microwave Kinetic Inductance Detectors using principal component analysis}

\author[a]{Jacob M. Miller}
\author[a]{Nicholas Zobrist}
\author[b]{Gerhard Ulbricht}
\author[a,*]{Benjamin A. Mazin}
\affil[a]{University of California, Department of Physics, Santa Barbara, CA, USA, 93106}
\affil[b]{Dublin Institute of Advanced Studies, School of Cosmic Physics, 31 Fitzwilliam Place, Dublin 2, D02XF86, Ireland}

\cftpagenumbersoff{figure}
\cftpagenumbersoff{table} 
\begin{document} 
\maketitle

\begin{abstract}
\hl{We develop a photon energy measurement scheme for single photon counting Microwave Kinetic Inductance Detectors (MKIDs) that uses principal component analysis (PCA) to measure the energy of an incident photon from the signal (``photon pulse'') generated by the detector. PCA can be used to characterize a photon pulse using an arbitrarily large number of features and therefore PCA-based energy measurement does not rely on the assumption of an energy-independent pulse shape that is made in standard filtering techniques. A PCA-based method for energy measurement is especially useful in applications where the detector is operating near its saturation energy and pulse shape varies strongly with photon energy.} It has been shown previously that PCA using two principal components can be used as an energy-measurement scheme. We extend upon these ideas and develop a method for measuring the energies of photons by characterizing their pulse shapes using any number of principal components and any number of calibration energies. Applying this technique with 50 principal components, we show improvements to a previously-reported energy resolution for \hl{Thermal Kinetic Inductance Detectors (TKIDs)} from 75 eV to 43 eV at 5.9 keV. We also apply this technique with 50 principal components to data from an optical to near-IR MKID and achieve energy resolutions that are consistent with the best results from existing analysis techniques.
\end{abstract}


{\noindent \footnotesize\textbf{*}Benjamin A. Mazin,  \linkable{bmazin@physics.ucsb.edu } }

\begin{spacing}{2}   

\section{Introduction}

Microwave Kinetic Inductance Detectors (MKIDs) are \hl{superconducting sensors}~\cite{Day2003, Ulbricht2015,Szypryt2017b} used for sensitive astronomical observations. These devices use changes in the surface impedance of a superconductor to sense individual photon impacts with up to microsecond precision. The superconductor is patterned into a microwave resonator which allows each sensor to be addressed at a different frequency on the same feedline. This multiplexing scheme dramatically simplifies the readout of the system compared to other superconducting detector technologies, and large arrays of up to 20,000 detectors have already been demonstrated~\cite{Meeker2018,Walter2020}.

An important quality in an MKID is the precision with which it can measure the energy of each incident photon --- its energy resolution. For MKIDs operating in the optical wavelength range, improvements to the energy resolution open the way for accurate spectral measurements~\cite{Rauscher2016} and observation at higher contrast ratios~\cite{Wang2017} in the \hl{direct imaging of exoplanets}. In the X-ray regime, thermal KIDs (TKIDs) have been proposed as an alternative to the more sensitive but harder to multiplex transition edge sensors (TESs). Closing the gap between the resolution of TKIDs and TESs in this energy range is important for applications where both high spatial and spectral resolution is needed~\cite{Barcons2015}.

Improvements to energy resolution for these types of detectors occur in two domains: reduction of noise in hardware and data-processing in software. While efforts in both domains are readily pursued, this work will focus on the \hl{latter}. The most widely-used technique for measuring the energy of a photon pulse is called \hl{optimal filtering}~\cite{McCammon1988, Fowler2016}. In this method, a digital filter is constructed that extracts the energy of each photon by modeling the detector response as a pulse with a fixed shape and a variable amplitude. The photon energy is related to the pulse amplitude by calibrating the detector with photons of known energy. With this pulse model and filter, the energy estimated for each photon has the best possible energy resolution under the constraint of fixed pulse shape~\cite{Fowler2015}. However, in real detectors this assumption discards additional information about the energy contained in the shape and can increase the energy uncertainty, degrading the energy resolution~\cite{Fowler2017}.  

To create an energy-fitting model that accounts for variations in shape, the use of statistical feature-extraction tools such as principal component analysis (PCA) have been proposed~\cite{Yan2016, Busch2016}. PCA can be used on existing data to determine the most strongly-varying features of pulse shape with energy. The presence of these features in future pulses can then be used to predict the energy of the pulse. In this way, it is possible to create a model for energy that is easily calibrated on existing data and that utilizes information contained in the shape of each pulse, which is no longer assumed to be fixed.

In previous works, PCA methods for photon detector analysis have been applied to TKID data~\cite{Yan2016} as well as TES data~\cite{Busch2016}. Both studies demonstrated that representing the data in a two-dimensional subspace formed by two principal components is an effective way to distinguish photons with two distinct energies and improve the energy resolution. Additionally, a similar algorithm has been employed to \hl{identify and reject near-coincident events} in detectors that have significant pulse shape variation~\cite{Alpert2016}. In all of these cases, the PCA methods were shown to be more robust estimators than optimal filtering.

\hl{We begin in section}~\ref{sec:pca} \hl{by reintroducing the fundamentals of the PCA method, and in section}~\ref{sec:tkid} \hl{we reanalyze TKID data from reference} \citenum{Ulbricht2015} \hl{and build upon the existing literature by showing that there can be significant advantages to including more than the first two PCA components.} In section~\ref{sec:opt}, \hl{we develop} techniques to calibrate such a model on data containing photons from more than two distinct source energies, allowing us to apply these methods to optical MKIDs. \hl{For each analysis, we pre-process the data to reject coincident photon events and align photon pulse start times.}

\section{Principal Component Analysis} \label{sec:pca}

\hl{A photon pulse is a signal from a photon detection that consists of a sequence of $M$ voltage measurements} taken at times relative to the photon arrival time. To calibrate our energy-fitting model discussed in sections~\ref{sec:tkid} and~\ref{sec:opt}, we need to process a set of $N$ pulses (up to $N \sim 10^4$) without discarding any potentially useful information. However, the typically large dimensionality of each pulse (up to $M \sim 10^4$) and the nature of our calibration technique makes it computationally unrealistic to use the full pulse. Instead, we use principal component analysis (PCA) to find a low-dimensional basis space into which we can project our pulses before running calibration.
The idea behind PCA is to construct an ordered basis such that the first basis vector (or ``principal component") is the direction of largest variance in the data and each following basis vector is the direction of next largest variance that is also orthogonal to the previous directions. We compute this basis for a set of photon pulses containing measurements of two or more distinct energies, which introduces a strong source of energy-dependent pulse shape variation. By ensuring that the variance in pulse shape between distinct energies is large compared to the variance in pulse shape due to other sources, we can expect to capture the majority of information about photon energy in a subspace including only the first several principal components. We can then project our pulses into this subspace to reduce their dimensionality while preserving information about photon energy and rejecting extraneous information including noise.


 The unordered principal components can be calculated by writing the covariance matrix for our data and computing its eigenbasis~\cite{Bishop2006}. For a set of $N$ pulses $\mathbf{x}_1$, $\mathbf{x}_2$, ..., $\mathbf{x}_N$ with $M$ measurments each, we collect the data into a matrix $\mathbf{\tilde{X}}_\text{M x N}$ with columns containing mean-centered pulses $\mathbf{\tilde{x}}_1$, $\mathbf{\tilde{x}}_2$, ..., $\mathbf{\tilde{x}}_N$ which are defined such that $\mathbf{\tilde{x}}_i = \mathbf{x}_i - \mathbf{\bar{x}}$ where $\mathbf{\bar{x}} = \frac{1}{N} \sum_{i=1}^N {\mathbf{x}_i}$. The covariance matrix is then defined as
\begin{equation}
    \label{eq:cov}
    \mathbf{C}_\text{M x M} = \frac{1}{N-1} {\mathbf{\tilde{X}}}_\text{M x N} {\mathbf{\tilde{X}}^{\rm T}}_\text{N x M}.
\end{equation}
\sloppy Diagonalizing this covariance matrix results in eigenvectors $\mathbf{u}_1, \mathbf{u}_2, ...,\mathbf{u}_M$ and eigenvalues ${\sigma^2_1, \sigma^2_2, \dots, \sigma^2_M}$, which can be shown to equal the variances of the data projected on to each corresponding eigenvector~\cite{Bishop2006}. The ordered basis of principal components is constructed by sorting the eigenvectors by descending variance.


In practice we can avoid the computation that is required to construct and diagonalize the covariance matrix $\mathbf{C}_\text{M x M}$ by deriving its eigenbasis directly from $\mathbf{\tilde{X}}_\text{M x N}$ using singular value decomposition (SVD)~\cite{Strang2006}. The SVD factors $\mathbf{\tilde{X}}_\text{M x N}$ so that
\begin{equation}
    \label{eq:svd}
    \mathbf{\tilde{X}}_\text{M x N} = \mathbf{U}_\text{M x M} \mathbf{\Sigma}_\text{M x N} {\mathbf{V}^{\rm T}}_\text{N x N},
\end{equation}
where $\mathbf{U}_\text{M x M}$ and ${\mathbf{V}^{\rm T}}_\text{N x N}$ are orthogonal, and $\mathbf{\Sigma}_\text{M x N}$ is diagonal in its left $M$~x~$M$ square (if $M<N$) or upper $N$~x~$N$ square (if $N<M$) and zero everywhere else. The SVD exists for any matrix~\cite{Strang2006} and can be computed using most linear algebra programming packages. By substituting Eq~\ref{eq:svd} in into Eq~\ref{eq:cov}, we can write
\begin{align}
\begin{split}
    \mathbf{C}_\text{M x M} & = \frac{1}{N-1} (\mathbf{U}_\text{M x M} \mathbf{\Sigma}_\text{M x N} {\mathbf{V}^{\rm T}}_\text{N x N})({\mathbf{V}}_\text{N x N} {\mathbf{\Sigma}^{\rm T}}_\text{N x M}{\mathbf{U}^{\rm T}}_\text{M x M})\\
    & = \frac{1}{N-1}\mathbf{U}_\text{M x M} \mathbf{\Sigma}_\text{M x N} {\mathbf{\Sigma}^{\rm T}}_\text{N x M}{\mathbf{U}^{\rm T}}_\text{M x M}\\
    & = \frac{1}{N-1}\mathbf{U}_\text{M x M} \mathbf{\Sigma^2}_\text{M x M} {\mathbf{U}^{\rm T}}_\text{M x M},
\end{split}
\end{align}
which is the eigendecomposition of $\mathbf{C}_\text{M x M}$ up to a constant factor. Therefore, the columns of the matrix $\mathbf{U}_\text{M x M}$ of the SVD give us the principal components $\mathbf{u}_1, \mathbf{u}_2, ...,\mathbf{u}_M$ for our data. The square roots of the corresponding variances $\sqrt{\sigma^2_1}$, $\sqrt{\sigma^2_2}$, ..., $\sqrt{\sigma^2_M}$ are also given by the SVD (up to a constant factor) as the \hl{non-zero} elements of $\mathbf{\Sigma}_\text{M x N}$. The principal components must then be sorted by decreasing variance, which is done automatically by most packaged SVD implementations.

Dimensionality reduction of our data from dimension $M$ to some lower dimension $K$ is performed by projecting $\mathbf{x}_1$, $\mathbf{x}_2$, ..., $\mathbf{x}_N$ into the subspace of the first $K$ ordered principal components. Previous works have explored pulse analysis methods in a subspace limited to two principal component dimensions\cite{Yan2016, Busch2016}. In our work, we explore extending these methods to an arbitrary dimensionality $K$ to improve the resolution of the detector. \hl{Importantly, PCA allows us to select $K<<M$ while keeping the fraction of variance that is accounted for by the reduced basis, $\sum_{i=1}^{K} {\sigma_i'}^2 / \sum_{i=1}^{M} {\sigma_i'}^2$,
significantly close to one.} Here, $\sigma_i'$ is used instead of $\sigma_i$ to signify that the variances are reordered from largest to smallest. The exact choice of $K$ depends on the underlying distribution of the data, the nature of any noise present, and the cost of dimensionality in the data-processing techniques to be used.

\section{TKID Data Analysis} \label{sec:tkid}

\begin{figure}
    \centering
    \includegraphics[height=3in]{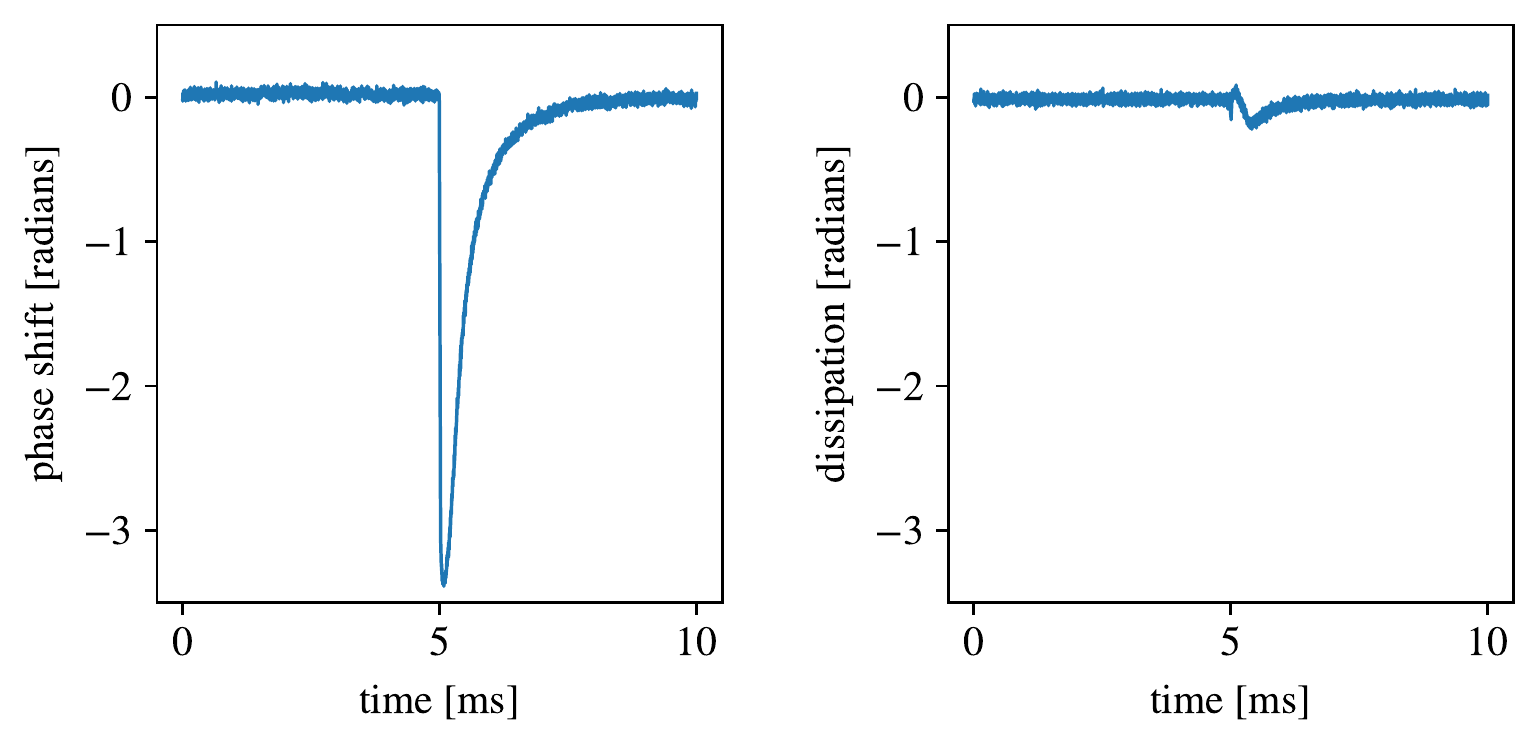}
    \caption{Photon pulses in our data set contain both a phase and dissipation measurement, each containing 8,000 samples over the span of 10 ms.}
    \label{fig:trace}
\end{figure}

To test the effect of increasing PCA dimension on energy resolution we use data \hl{from reference} \citeonline{Ulbricht2015} \hl{that is measured} with a TKID illuminated by an \hl{iron} source, which has X-ray emission line energies within the TKID’s dynamic range at approximately 5.89 keV, 5.90 keV, and 6.49 keV. The first and second lines are $\approx 60$x closer to each other than they are to the third line, so we focus on resolving only two separate energy peaks. \hl{Each photon pulse measurement from a kinetic inductance detector includes a pulse related to the phase of the signal and a pulse related to the power, or dissipation, of the signal.} The data set we use contains $N=3,552$ distinct 10 ms photon pulses with $8,000$ phase measurements and $8,000$ dissipation measurements each. Examples of the phase and dissipation pulses for a photon event are shown in Fig~\ref{fig:trace}.
For the $i$th photon event, the phase and dissipation data are joined end-to-end to create the $M=16,000$ component data trace $\mathbf{x}_i$ that is used in PCA. \hl{The phase and dissipation pulses are joined without any relative scaling to ensure that the noise floor of each pulse is matched. In this way, the variance in the combined trace due to noise from either pulse will be minimized relative to the variance in the combined trace due to the signal in either pulse. This enforces an ordering of principal components where components related to the phase and dissipation signal are sorted to the beginning and components related to the noise in either pulse are sorted to the end.}

\begin{figure}
    \centering
    \includegraphics[height=3.5in]{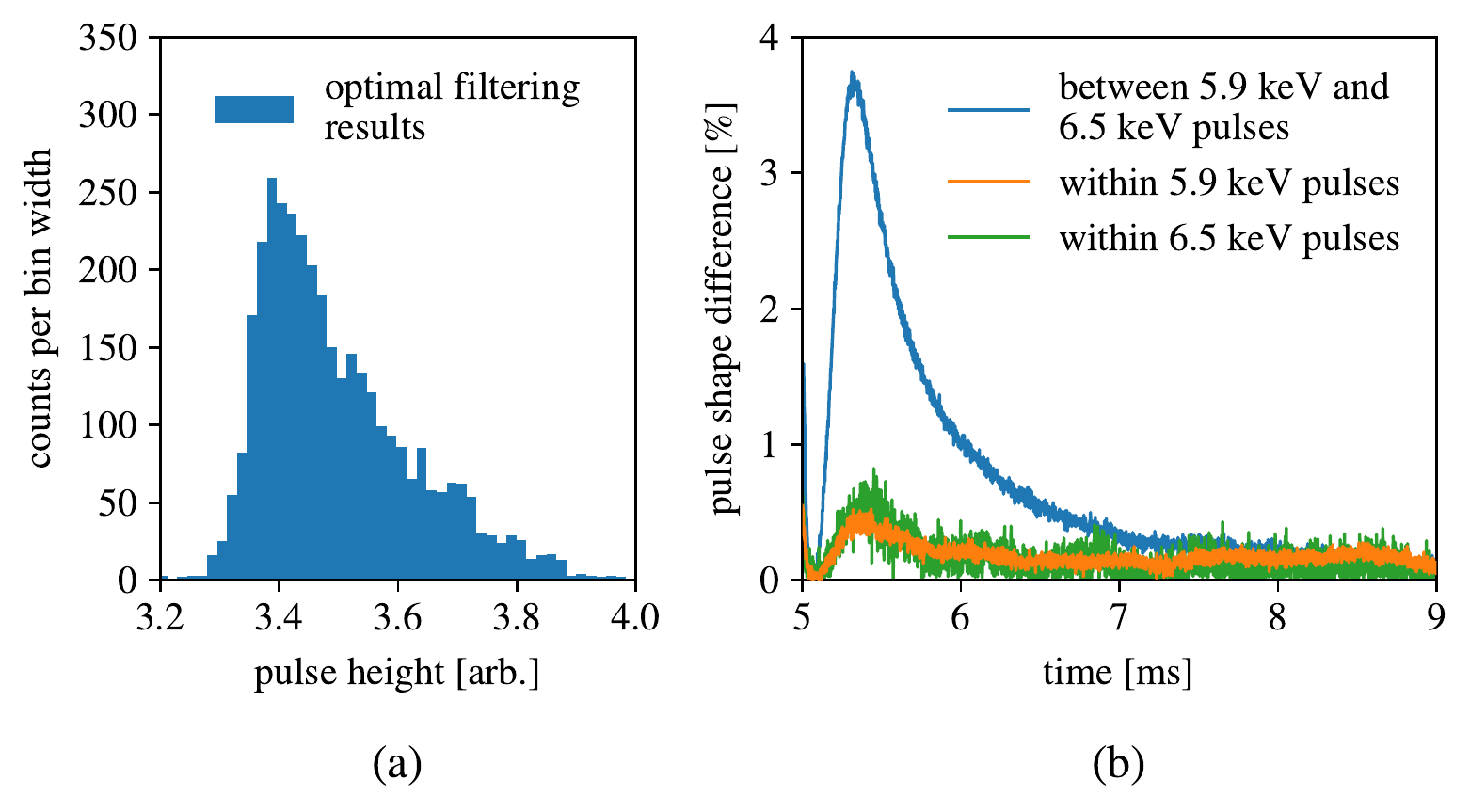}
    \caption{\hl{The distribution of pulse heights as calculated using optimal filtering on the phase and dissipation signal is plotted in \textbf{(a)}. The two source energies cannot be resolved using optimal filtering because pulse shape is varying strongly with energy. The shape variation between phase signal pulses is characterized in \textbf{(b)}. Shape difference is computed between two pulses by normalizing each pulse to its height and computing the difference. Using the results of section}~\ref{sec:tkid}\hl{, we can predict the photon energy for each pulse with high enough resolution that the resulting energy distribution contains two finite-width peaks corresponding to 5.9 keV and 6.5 keV photons (Fig}~\ref{fig:tkidbest}\hl{). The blue line in \textbf{(b)} shows the shape difference between the average pulses from each peak in this energy distribution. The orange and green lines show the shape difference between the average pulses from the bottom 25\% and top 25\% of the 5.9 keV and 6.5 keV peaks respectively.}}
    \label{fig:nonlin}
\end{figure}

\hl{Optimal filtering techniques used on this data fail to resolve the different photon source line energies as shown in the left frame of Fig}~\ref{fig:nonlin}.
\hl{This poor performance occurs because the detector is measuring photons at energies which are high enough that the height of the resulting photon pulse becomes saturated and the pulse shape becomes dependent on energy. The degree of shape variation is plotted in the right frame of Fig~}\ref{fig:nonlin} \hl{for pulses between each source energy and for pulses within each source energy. We see that the majority of pulse shape variation in this data occurs between source energies while energy-independent pulse shape variations are comparatively small. For pulses with saturated height but energy-dependent pulse shape variation, the energy of an incident photon can be predicted using features of the pulse shape in addition to pulse height.} We expect that PCA will successfully isolate the features that are most strongly related to changing energy because we have configured the calibration data such that pulse shape variations between line energies comprise the largest source of variance. Each of the pulses can then be projected on to these features, which are the principal components, before performing analysis to determine their energies.

\begin{figure}
    \centering
    \includegraphics[height=3.5in]{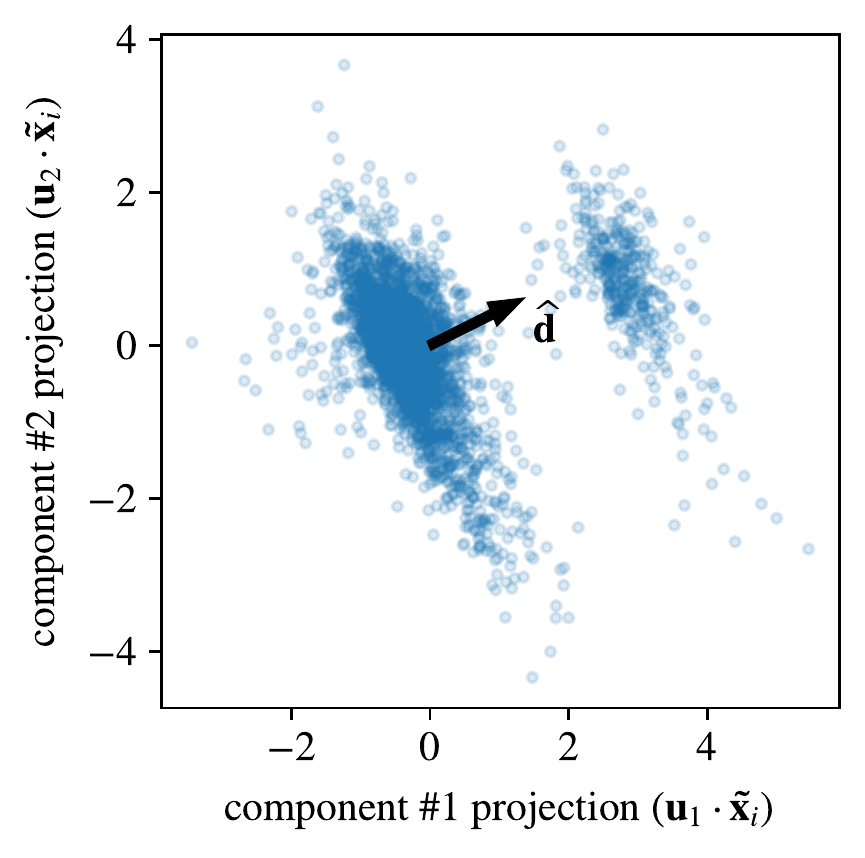}
    \caption{Projecting the photon pulses into the basis space of their first two principal components reveals underlying trends in the data. In this space, variance due to energy appears as a separation of the pulses into two distinct clusters. The direction $\mathbf{\widehat{d}}$ represents the direction of changing energy in this space.}
    \label{fig:scatter}
\end{figure}

Similar to results in previous works, the two-dimensional PCA projection of the TKID data plotted in Fig~\ref{fig:scatter}
separates the photon pulses into two clusters which represent the two energy peaks present in our data. This separation occurs because the variation in pulse shape due to changing energy is much larger than the variation in the pulse shape due to noise and is therefore strongly captured using only the first two principal components. After our pulses have been projected into the $K$-dimensional subspace ($K=2$ in Fig~\ref{fig:scatter}), we assign an energy to each pulse by searching for the unit direction of changing energy $\mathbf{\widehat{d}}$ and projecting again onto this direction. If $\mathbf{\tilde{U}}_\text{M x K}$ is defined as the matrix containing the first $K$ out of $M$ principal components in its columns, then ${\mathbf{\tilde{U}}^{\rm T}}_\text{K x M} \; \mathbf{\tilde{x}}_i$ is the projection of $\mathbf{\tilde{x}}_i$ into the subspace of the first $K$ principal components. The energy $E_i$ of the $i$th pulse can be written as a function $f(x)$ of this $K$-dimensional pulse projected on to $\mathbf{\widehat{d}}$:
\begin{equation}
    E_i=f \bigg ( \; \big ({\mathbf{\tilde{U}}^{\rm T}}_\text{K x M} \; \mathbf{\tilde{x}}_i \big ) \cdot \mathbf{\widehat{d} \; \bigg )}.
    \label{eq:trans}
\end{equation}
Here we calibrate the detector by approximating $f(x)$ as a linear transformation $f(x) = C_1 x + C_0$ with constants $C_1$ and $C_0$ determined by aligning the median of both peaks of the projected distribution $\{E_i\}$ with the known line energies. We align distributions using the median rather than the mean or mode because the mean is overly sensitive to skew and because estimating the mode from a smoothed histogram is slow and error-prone for irregularly-shaped histograms.

We determine $\mathbf{\widehat{d}}$ by finding the direction in our subspace that produces an energy distribution $\{E_i\}$ with the narrowest peaks. Our data contains measurements from line emission sources, so a higher resolution measurement produces a distribution with peaks approaching delta functions. \hl{The Shannon entropy~}\cite{Shannon1948} \hl{has been used in several fields as a metric for the presence of sharp features in a distribution or spectrum}~\cite{Fowler2016, Misra2004}. We define the \hl{Shannon entropy} as
\begin{equation}
    \label{eq:ent}
    \mathrm{H} = -\sum_{i=1}^n \mathrm{P}(\varepsilon_i) \ln{\mathrm{P}(\varepsilon_i)}
\end{equation}
where the energy $\varepsilon_i$ is the $i$th outcome of a simulated a discrete random variable that is defined by binning $\{E_i\}$ into $n$ bins of fixed width and the probability $\mathrm{P}(\varepsilon_i)$ is the fraction of total pulses in the $i$th bin. We choose a fixed bin width that is small to produce a probability distribution that approximates a continuous curve while \hl{ensuring} it is not so small that our distribution is affected by empty bins. Results are however not highly sensitive to the exact value of bin width. We can then find the best $\mathbf{\widehat{d}}$ by searching for the direction that minimizes $H$.



Optimizing $\mathbf{\widehat{d}}$ can be done using a number of methods depending on whether speed or accuracy is important. A $K$-dimensional ``full" optimization can be performed by parameterizing the direction $\mathbf{\widehat{d}}$ in terms of $K-1$ angles using generalized spherical coordinates. However, the computational complexity increases quickly with $K$. To avoid slow high-dimensional optimization, it is also possible to use a recursive approach which is faster but does not search the full parameter space. \hl{The recursive approach starts by finding the direction of changing energy $\mathbf{\widehat{d}}$ in the space of the first two principal components by sweeping one polar angle. The algorithm then searches for $\mathbf{\widehat{d}}$ in three dimensions by starting with the optimum direction in two dimensions and performing a one-dimensional search along the direction of the third principal component. The fourth through $K$th principal component directions are added and optimized one-by-one until an approximate value for $\mathbf{\widehat{d}}$ in $K$ dimensions is found.} It is important that the specific optimization routine used is stochastic (does not use gradients) because Shannon entropy as a function of $\mathbf{\widehat{d}}$ is discontinuous at a small scale. This discontinuity arises because we compute Shannon entropy from a histogram with finite bin size and the value only changes when $\mathbf{\widehat{d}}$ moves enough to shift a pulse across the border between bins. The optimization routine we use \hl{is \texttt{differential\_evolution} from \texttt{scipy.optimize}}~\cite{2020SciPy-NMeth}.

\begin{figure}
    \centering
    \includegraphics[height=3in]{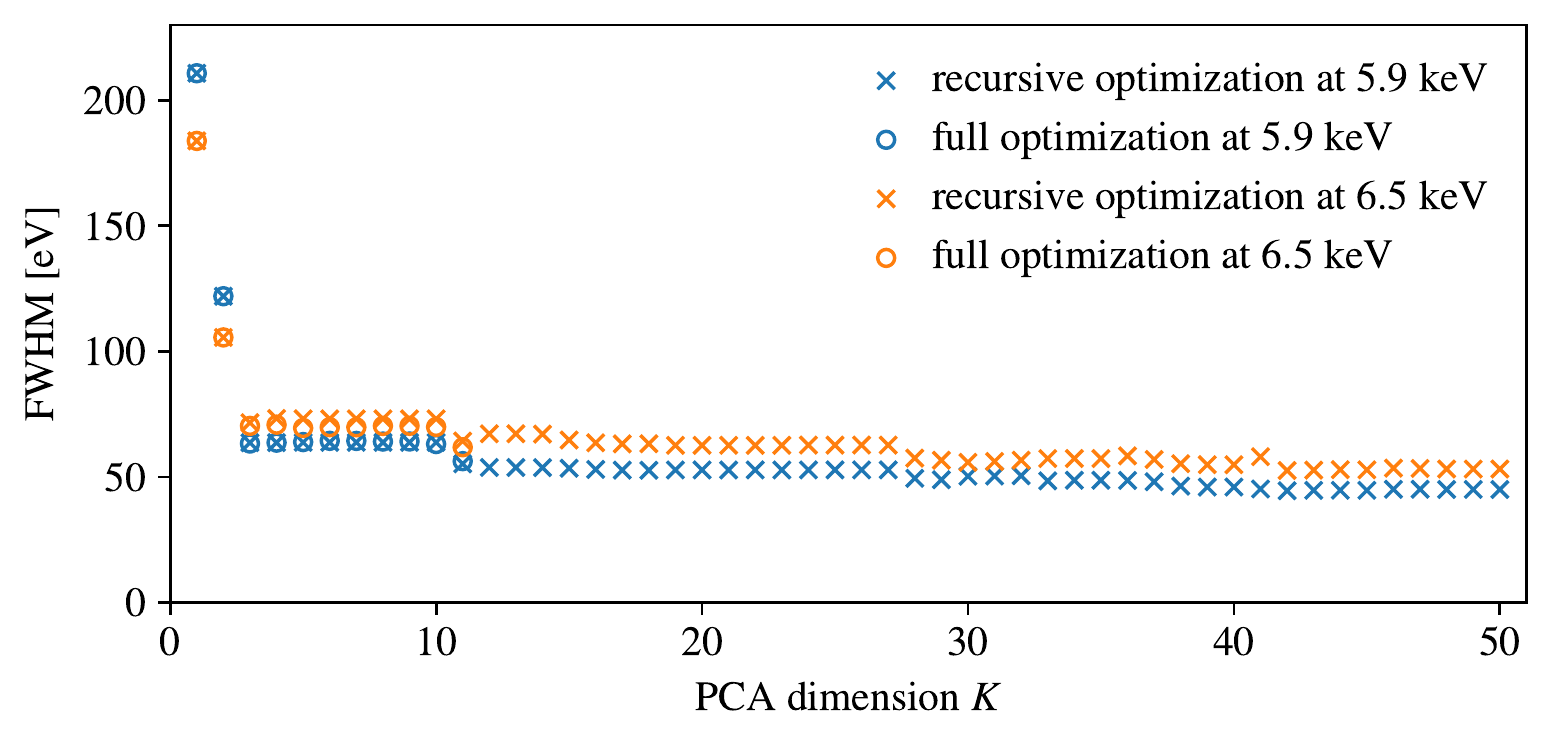}
    \caption{The energy resolution of both the 5.9 keV and 6.5 keV peaks improves as more PCA dimensions are used. The full optimization is computationally intensive for increasing dimension and is shown only for $K \le 11$. However, the recursive optimization yields comparable results and can be performed for large $K$. It is important to note that we plot FWHM but we are optimizing entropy and that the two quantities are not strictly positively correlated. Although the full optimization always offers a lower entropy than the recursive optimization and entropy strictly decreases with each additional dimension, these trends will not necessarily be true of the FWHM.}
    \label{fig:compcontrib}
\end{figure}

The result of a detector's energy resolution at some energy is typically reported using the full width half maximum (FWHM) of a measured line emission at this energy where a lower FWHM means better resolution. Figure~\ref{fig:compcontrib}
shows the computed FWHM of the 5.9 keV and 6.5 keV peaks as a function of increasing dimension when using both the full and recursive optimization method. This plot shows that we can improve on the results of two dimensional PCA significantly by increasing the dimensionality $K$ of our PCA space. The plot also demonstrates that using a recursive optimization technique does not harm the resolution of our analysis.

\begin{figure}
    \centering
    \includegraphics[height=3.5in]{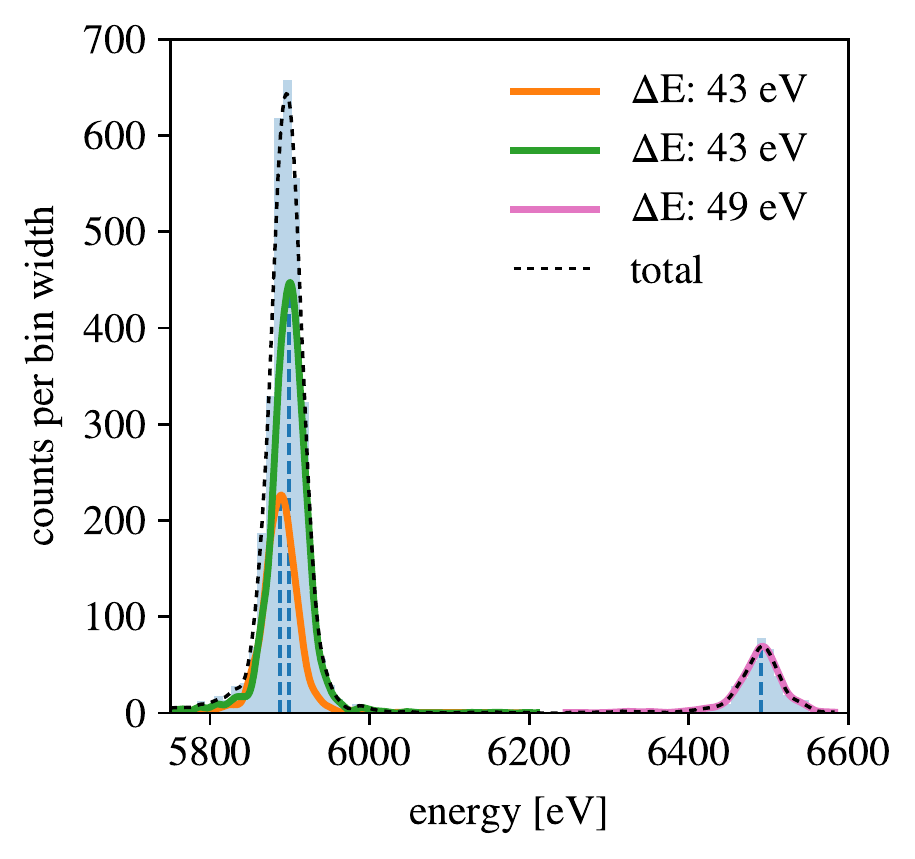}
    \caption{Using $K=50$ dimensions we achieve a FWHM (labeled $\Delta$E in the legend) of 43 eV for the two peaks at 5.89 keV and 5.90 keV and 49 eV for the peak at 6.49 keV. These values improve over the results using $K=2$ dimensions by 3x for the lower peaks and 2x for the upper peak. The dashed vertical lines are located at each known emission energy and the median of each distribution falls on these lines.}
    \label{fig:tkidbest}
\end{figure}

The results for this method using recursive optimization with $K=50$ are presented in Fig~\ref{fig:tkidbest}
and demonstrate a FWHM of 43 eV for the two lower peaks and a FWHM of 49 eV for the upper peak. Compared to results for this method using only $K=2$, this is approximately a 3x improvement for the lower peaks and a 2x improvement for the upper peak. Our results also improve upon the resolution achieved for this detector by Ulbricht et al. who demonstrated a FWHM of 75 eV at the lower peak by modeling the detector response and using curve fitting techniques\cite{Ulbricht2015}.


\section{Optical to Near-IR MKID Data Analysis} \label{sec:opt}

The PCA pulse analysis technique discussed in section \ref{sec:tkid} finds the direction of changing energy $\mathbf{\widehat{d}}$ between exactly two calibration energies. By using only two calibration points, we assume that every pulse has an energy that depends linearly on its projection onto $\mathbf{\widehat{d}}$ and thus choose a linear transformation for $f(x)$ in Eq~\ref{eq:trans}. However, if we have more than two calibration energies available, we can use them to find a nonlinear transformation $f(x)$ that takes as an input a pulse's projection onto $\mathbf{\widehat{d}}$ and outputs the corresponding photon's energy.

The data we use to develop a technique for multi-peak calibration is measured by reference \citeonline{Zobrist2021} and comes from an optical to near-infrared (near-IR) MKID illuminated by seven laser sources at energies of 0.94 eV, 1.11 eV, 1.26 eV, 1.35 eV, 1.52 eV, 1.87 eV, and 3.05 eV. The data contains between $N=4,000$ and $N=9,000$ distinct 1 ms photon pulses for each energy and $M=5,000$ samples for each pulse with $2,500$ from phase measurement and $2,500$ from dissipation.

\begin{figure}
    \centering
    \includegraphics[height=3.5in]{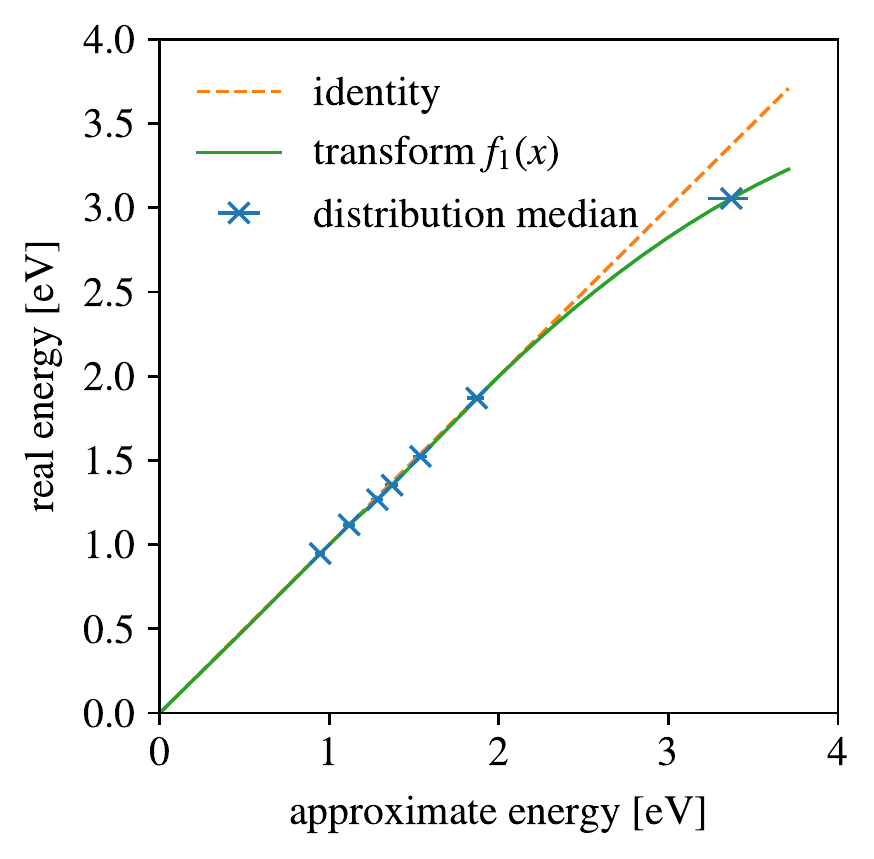}
    \caption{The real energy of each distribution is plotted as a function of that distribution's approximate energy as determined by a linear calibration using the 0.94 eV (first) and 1.87 eV (sixth) peak. These peaks were chosen because they lie at either end of the main cluster of energies. The transform $f_1(x)$ is a second-order spline generated with ``not-a-knot" boundary conditions that passes through (0, 0). \hl{The horizontal error bar on each median spans the middle 50\% of the corresponding distribution.}}
    \label{fig:transres}
\end{figure}

Our approach when calibrating with multiple energies is to begin with a linear guess $f_0(x)$ for $f(x)$ and to iteratively converge on a nonlinear solution. For $f_0(x)$, we choose a pair of energies from the source and find $\mathbf{\widehat{d}}$, $C_0$, and $C_1$ as we did in section \ref{sec:tkid} by minimizing the distribution's entropy and properly aligning the resulting distribution with the \hl{pair of known line energies}. We note that we now minimize using the joint entropy which is the average of the entropy (as defined in Eq~\ref{eq:ent}) of each energy distribution weighted by the number of samples in each distribution. We previously used the total entropy of the combined distribution, but using joint entropy is possible for the optical data because each energy was measured separately and thus the pulses can be labeled. If we plot the real energy as a function of the energy approximated by $f_0(x)$, we find the relationship plotted in Fig~\ref{fig:transres}.
This figure shows that our initial guess loses accuracy at higher photon energies where the detector is saturating, far from the initial calibration energies.

We can improve upon our initial guess by smoothly connecting the points in Fig~\ref{fig:transres} to build a new function $f_1(x)$ that transforms the approximate energies as computed by $f_0(x)$ onto the real energies. We then set $f(x) = f_1(f_0(x)) \equiv f_1 \circ f_0(x)$ in Eq~\ref{eq:trans} and re-optimize $\mathbf{\widehat{d}}$ to find the best direction in the new energy space, which is closer to the real one. Re-optimization is necessary because the new $f(x)$ will return a different distribution $\{E_i\}$ as function of $\mathbf{\widehat{d}}$ (see Eq~\ref{eq:trans}). This means that the best $\mathbf{\widehat{d}}$ may also change and needs to be recomputed.

This iteration can be continued indefinitely by computing a new transform $f_i(x)$ between the approximate energies of the previous iteration and the real energies and repeating the optimization using $f(x) = f_i \circ f_{i-1} \circ \cdots \circ f_1 \circ f_0(x)$. In practice however, iterating past $f_1(x)$ is unnecessary because the median of each distribution resulting from $f(x) = f_1 \circ f_0(x)$ for this data is nearly equal to the real energy.

\begin{figure}
    \centering
    \includegraphics[height=3in]{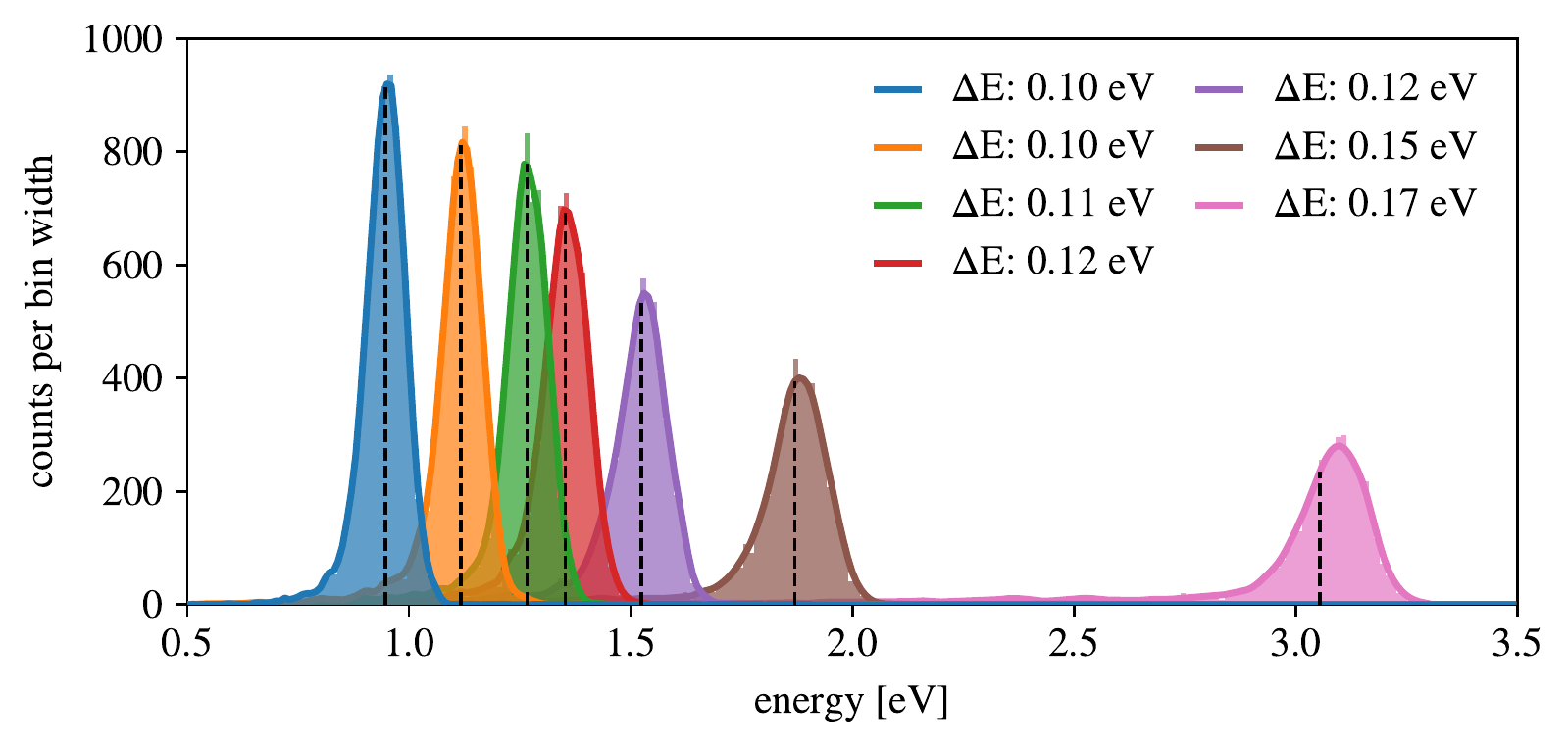}
    \caption{Using $K=50$ dimensions we achieve energy resolutions (FWHM) of 0.10~eV, 0.10~eV, 0.11~eV, 0.12~eV, 0.12~eV, 0.15~eV, and 0.17~eV for photons with respective energies of 0.94~eV, 1.11~eV, 1.26~eV, 1.35~eV, 1.52~eV, 1.87~eV, and 3.05~eV. The results in this plot are achieved using two iteration of optimization. The dashed vertical lines are located at each known emission energy and the median of each distribution falls on these lines. We do not discard the long tails of the upper distributions and instead allow them to pull the median. Because each energy was measured independently, we trust that the pulses in the distribution tails are labeled properly and want our calibration to incorporate these pulse shapes as best as possible.
    }
    \label{fig:optres}
\end{figure}

Figure~\ref{fig:optres}
shows the results of the multi-peak PCA method using 50 components and two rounds of optimization. These results are consistent with the energy resolutions achieved in reference \citenum{Zobrist2021} by using filtering on the same data. The inability of the feature extraction technique to improve upon the these results suggests that the energy resolution of this data is limited by variations in pulse shape that are independent of energy. Some suspected sources of these energy-independent shape changes are phonon loss into the substrate~\cite{Fyhrie2020} and a position-dependent detector response~\cite{deVisser2021,Zobrist2019}. Further improvements to the detectors themselves are likely required before the full benefits of using this analysis technique can be realized.

\section{Conclusion} \label{sec:conc}
Filtering as an energy measurement technique performs poorly when used with data containing energy-dependent pulse shape variations because the technique relies on an assumption that the shape is constant. One common example of energy-dependent shape variation is data from a detector that is operating near its saturation energy, such as the TKID data discussed in section~\ref{sec:tkid}. For such cases of non-constant pulse shape, energy resolution can be improved by analyzing how pulse shape varies with energy. In this work we extended upon previous ideas to develop a technique for energy measurement that uses PCA to incorporate information contained in the variation of the pulse shape with energy. We characterized the energy dependence of the pulse shape using up to 50 principal components and found that it is beneficial to include more than two components \hl{when detector resolution is impaired by energy-dependent changes in pulse shape}. We also showed one way to calibrate the energy measurement using data from more than two \hl{source energies}.

Using PCA-based energy measurement, \hl{we demonstrated an improved energy resolution of nearly twice the previous best results} for a saturated TKID in the X-ray regime. This is consistent with our hypothesis that we can improve energy resolution in situations of non-constant pulse shape by analyzing the energy dependence of the dominant features of the pulse. We also measured the energy resolution of optical to near-IR KID data and achieved results consistent with filtering results for the same data. The equivalence of these two methods on this detector corroborates our understanding that the energy resolution is limited by detector physics and not by the specific data analysis technique.

\subsection* {Acknowledgments}
Graduate student N.Z. was supported throughout this work by a NASA Space Technology Research Fellowship. Undergraduate student J.M. was supported throughout this work by the Eddleman Fellowship granted by the Eddleman Center for Quantum Innovation at UC Santa Barbara. This research was carried out in part at the Jet Propulsion Laboratory, under a contract with the National Aeronautics and Space Administration.

\nolinenumbers

\bibliography{report}   
\bibliographystyle{spiejour}   

\end{spacing}
\end{document}